\documentclass[12pt,preprint]{aastex631}
\usepackage{amsmath}
\usepackage{graphicx}
\usepackage{threeparttable}
\usepackage{txfonts}
\usepackage{soul}
\usepackage{newtxtext,newtxmath}

\usepackage[T1]{fontenc}

\begin{document}
\title{The physical origin of the periodic activity for FRB 20180916B}

\correspondingauthor{Fa-Yin Wang}
\email{fayinwang@nju.edu.cn}
\author{Hao-Tian Lan}
\affiliation{School of Astronomy and Space Science, Nanjing University Nanjing 210023, People's Republic of China}
\author[0000-0002-2171-9861]{Zhen-Yin Zhao}
\affiliation{School of Astronomy and Space Science, Nanjing University Nanjing 210023, People's Republic of China}

\author{Yu-Jia Wei}
\affiliation{Purple Mountain Observatory, Chinese Academy of Sciences, Nanjing 210023, China}

\author[0000-0003-4157-7714]{Fa-Yin Wang}
\affiliation{School of Astronomy and Space Science, Nanjing University Nanjing 210023, People's Republic of China}
\affiliation{Purple Mountain Observatory, Chinese Academy of Sciences, Nanjing 210023, China}
\affiliation{Key Laboratory of Modern Astronomy and Astrophysics (Nanjing University) Ministry of Education, People's Republic of
	China}
\begin{abstract} 
Fast radio bursts (FRBs) are transient radio signals with millisecond-duration, large dispersion measure (DM) and extremely high brightness temperature. Among them, FRB 20180916B has been found to have a 16-day periodically modulated activity. However, the physical origin of the periodicity is still a mystery. Here, we utilize the comprehensive observational data to diagnose the periodic models. We find that  the ultra-long rotation model is the most probable one for the periodic activity. However, this model cannot reproduce the observed rotation measure (RM) variations. We propose a self-consistent model, i.e., a massive star binary containing a slowly rotational neutron star and a massive star with large mass loss, which can naturally accommodate the wealth of observational features for FRB 20180916B. In this model, the RM variation is periodic, which can be tested by future observations.
\end{abstract}

\section{Introduction}

Fast radio bursts (FRBs) are transient radio signals which have a milliseconds-duration, large DM and extremely high brightness temperature \citep{Cordes2019,Xiao2021,Zhang2023}. They were first discovered in 2007 and then were verified in 2013 \citep{2007Sci...318..777L,2013Sci...341...53T}. Since then, hundreds of FRB sources had been detected and they are classified as repeating FRBs and non-repeating FRBs. Among the repeating FRBs, FRB 20180916B \citep{2020Natur.582..351C} and FRB 20121102 \citep{2020MNRAS.495.3551R,2021MNRAS.500..448C} are found to have a period of 16.35d and 157 d, which offers a great opportunity to study the physical origin of FRBs.
FRB 20180916B is a repeating FRB source first detected by the Canadian Hydrogen Intensity Mapping Experiment (CHIME) \citep{2019ApJ...885L..24C} and it was localized to a star-forming region in a nearby massive spiral galaxy with a redshift of $0.0337\pm0.0002$ \citep{2020Natur.577..190M}. Different from many other FRB sources, FRB 20180916B exhibits a period of 16.35 days and an active window of five days \citep{2020Natur.582..351C}. With the help of the follow-up observational data from the Low-Frequency Array (LOFAR) \citep{2021Natur.596..505P,Pleunis2021} and Apertif \citep{2021Natur.596..505P}, the active window shows a strong dependency on frequency, with higher frequency having a narrower active window and the phase center of higher frequency occurring earlier than the lower frequency \citep{2021Natur.596..505P}. For other observational properties, the DM of FRB 20180916B is approximately $349$ pc cm$^{-3}$ and its variation between different bursts is very small ($\Delta$DM $< 1$ pc cm$^{-3}$) \citep{2019ApJ...885L..24C,2023ApJ...950...12M,2021Natur.596..505P}. The RM of FRB 20180916B was stable \citep{2023ApJ...950...12M} in the first three years after its discovery in 2018. Then the RM increases monotonically over a 9 month duration by about $65.6$ rad m$^{-2}$ from $-120$ rad m$^{-2}$ to $-54.4$ rad m$^{-2}$. But the RM variations is unrelated to the 16.3-day period \citep{2023ApJ...950...12M}. For a single burst, the variation of linear polarization angle is less than $10^{\circ}\sim20^{\circ}$ and the bursts are $~100\%$ linearly polarised at high frequencies but depolarise towards low frequencies\citep{Nimmo2021,2024MNRAS.527.9872G}. For different bursts, the variation of linear polarization angle is less than $50^{\circ}$ \citep{Nimmo2021,Pleunis2021}. 

To explain this special periodic activity, numerous models have been proposed. Encouraged by the observation of FRB 20200428 which is associated with SGR J1935+2154 in the Milky Way \citep{CHIME/FRBCollaboration2020,Bochenek2020}, some theoretical models propose strongly magnetized neutron stars as the emitting source of FRBs. For these models, the periodic activity can be explained by the orbital motion of a binary system with a neutron star and a massive star with mass loss \citep{2020ApJ...893L..39L,2020ApJ...893L..26I,2021ApJ...920...54W,Li2021}, the free or forced precession of a neutron star \citep{2020ApJ...895L..30L,2020ApJ...892L..15Z,2020ApJ...893L..31Y,2020RAA....20..142T} or a ultra-long rotation of a neutron star \citep{2020MNRAS.496.3390B}. A binary system undergoing super-Eddington accretion can also be considered as the source of this FRB and the periodicity can be explained by the precession of the disc and the jet \citep{Katz2020,2021ApJ...917...13S}. These theoretical models not only make different predictions on the temporal evolution of the observed period \citep{2021MNRAS.502.4664K,2022A&A...658A.163W}, but also on the distribution of various observational parameters, such as DM, polarization angle and RM. There are also some other models, such as a magnetized pulsar traveling through an asteroid belt in a binary system \citep{2020ApJ...895L...1D}, a binary system with a neutron star and a white dwarf undergoing mass transfer \citep{2020MNRAS.497.1543G}, a planet partially disrupted by a neutron star at periastron \citep{Kurban2022} and a massive binary consisting of a magnetar and an early-type star \citep{Barkov2022}. Since they lack of explicit predictions, we will not discuss them in this work.

The paper is organized as follows. We calculate the period and period derivation of FRB 20180916B in Section~\ref{sec2}. Then, we use the observational data to constrain the periodic model in Section~\ref{sec3}. Finally, conclusions are given in Section~\ref{sec4}.

\section{Period and period derivation}\label{sec2}
Based on the discovery of periodically modulated activity \citep{2020Natur.582..351C}, the observational data of FRB 20180916B have significantly expanded which offers an opportunity to get a more precise period and its temporal evolution. Firstly, we calculate the period of FRB 20180916B with the epoch-folding method, which had been used to discover its period by CHIME team \citep{2020Natur.582..351C}. In this method, we first fold the arrival time of each burst into a normalization phase using the following equation
\begin{equation}
	\phi=\frac{t-t_0}{P},
\end{equation}
where $t_0$ is set as MJD $ = 58369.40$, $\phi$ is the normalization phase and P is the observational period. Then, the Pearson $\chi^2$ test is performed on the phase of the bursts against rectangular distribution for a given period. The higher $\chi^2$ of the period, the more reliability of the period.

We list the results in Table~\ref{table1}. In the first line of Table~\ref{table1}, we use the 122 bursts detected by CHIME up to the beginning of 2024 to calculate the period (https://www.chime-frb.ca/). The result is $16.33 \pm 0.04$ days, which is similar to the one found by CHIME/FRB \citep{2020Natur.582..351C}. We also utilize 29 bursts provided by LOFAR \citep{2021Natur.596..505P,Pleunis2021,2024MNRAS.527.9872G} and get a period of $16.32 \pm 0.06$ days, which is consistent with the result derived from the data of CHIME, as shown in the second line of Table~\ref{table1}. To investigate the possible temporal evolution of the period, we divide the bursts observed by CHIME into different time bins to calculate the period. In the last five lines of Table~\ref{table1}, we first divide the bursts of CHIME into two time bins. We get $16.33 \pm 0.10$ days and $16.32 \pm 0.07$ days which are simply the same. Then we divide the bursts of CHIME into three time bins. The results are $16.33 \pm 0.16$ days, $16.37 \pm 0.15$ days and $16.33 \pm 0.10$ days, respectively. These results support that the period of FRB 20180916B possesses a minimal variation, which falls well within the range of the error bar.
\begin{table*}
	
	\centering
	
	\caption{The period of FRB 20180916B derived from the epoch-folding method with use of the observation data of CHIME and LOFAR.}
	\begin{tabular}{cccccc}
		\hline
		Telescope & Burst number & Time range& Period (d) & Error (d) & References \\
		\hline
		CHIME     & 122   & 2018.09.16-2024.01.14   & 16.33  & 0.04  & https://www.chime-frb.ca/ \\
		LOFAR     & 29    & 2019.08.13-2022.09.28   & 16.32  & 0.06  & \citep{Pleunis2021,2024MNRAS.527.9872G} \\
		\hline
		CHIME     & 61    & 2018.09.16-2021-01-30   & 16.33  & 0.10  & \citep{2020Natur.582..351C} \\
		CHIME     & 61    & 2021-01-31-2024.01.14   & 16.32  & 0.07  & \citep{2023ApJ...950...12M} \\
		\hline
		CHIME     & 41    & 2018.09.16-2020.02.21   & 16.33  & 0.16  & \citep{2020Natur.582..351C} \\
		CHIME     & 41    & 2020.03.24-2021.09.30   & 16.37  & 0.15  & \citep{2023ApJ...950...12M} \\
		CHIME     & 40    & 2021.09.30-2024.01.14   & 16.33  & 0.10  & https://www.chime-frb.ca/ \\
		\hline
	\end{tabular}
	
	\label{table1}
	\vspace{0.5cm}
\end{table*}
In order to get a more accurate constraint on the period derivative ($\dot{P}$), we consider the effect of the $\dot{P}$ when calculating the phase in the epoch-folding method, i.e.,
\begin{equation}\label{eq1}
	\phi = \frac{2\pi (t-t_0)}{P}-\pi \frac{\dot{P}}{P^2}(t-t_0)^2.
\end{equation}
Two methods are used to estimate $P$ and $\dot{P}$. 
For the First method, we utilize the epoch-folding method as shown in Equation~(\ref{eq1}). The two-dimensional Gaussian functions are used to fit the result. The 1$\sigma$ range is taken as the error bar \citep{Sand2023}. We find $P = 16.31 \pm 0.32$ days and $\dot{P} = -2.6 \times 10^{-5} \pm 1.2 \times 10^{-4}$ day day$^{-1}$, which is consistent with the derivative $\dot{P}=0$. A similar result $\dot{P} = -2.0 \times 10^{-5} \pm 1.5 \times 10^{-4}$ is also found by CHIME team \citep{Sand2023}. However, it may be not appropriate to assume that the $\chi^2$ distributions of $P$ and $\dot{P}$ obeys two-dimensional Gaussian function. So we use the second method to give a $\dot{P}$ ranges for different periods.
\begin{figure}
	\centering
	\includegraphics[width=0.5\textwidth]{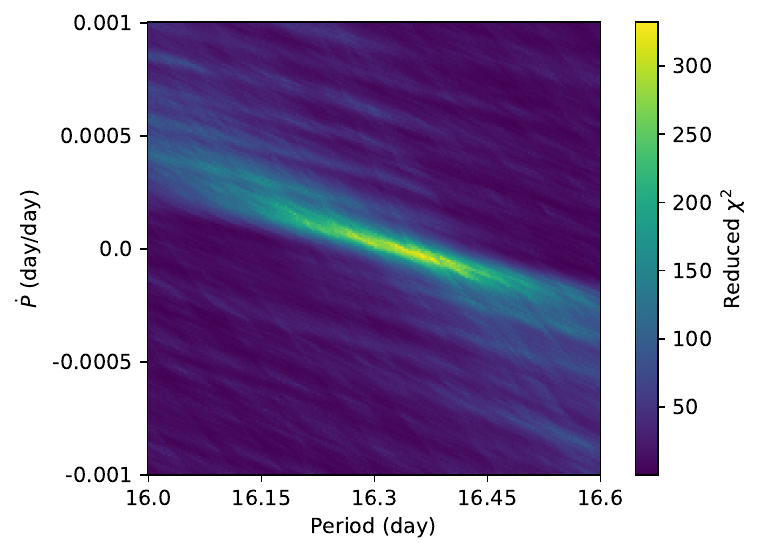}
	\includegraphics[width=0.5\textwidth]{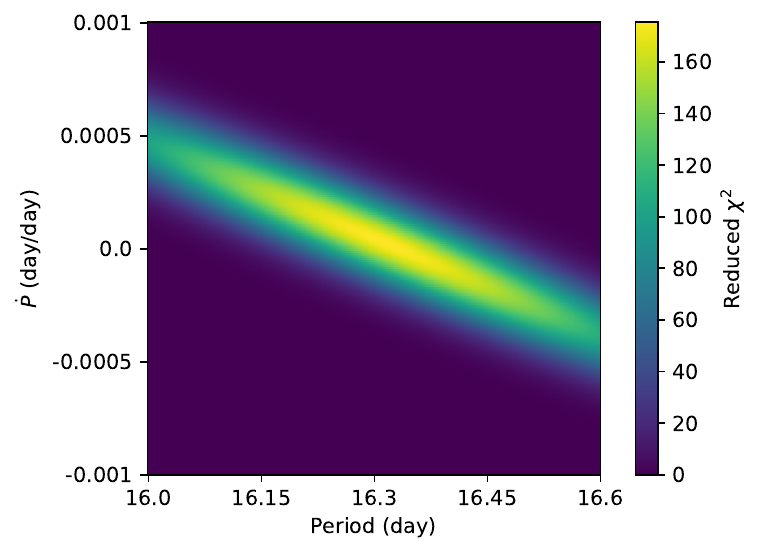}
	\caption{The reduced $\chi^2$ of different $P$ and $\dot{P}$. Top panel shows the result of epoch-folding method for different $P$ and $\dot{P}$ with the use of Equation~(\ref{eq1}). A higher $\chi^2$ means that the period and its $\dot{P}$ are more reliable respectively. Bottom panel is the two dimensional Gaussian fitting results of the top panel.}
	\label{fig4}
\end{figure}

\begin{figure}
	\centering
	\includegraphics[width=0.7\columnwidth]{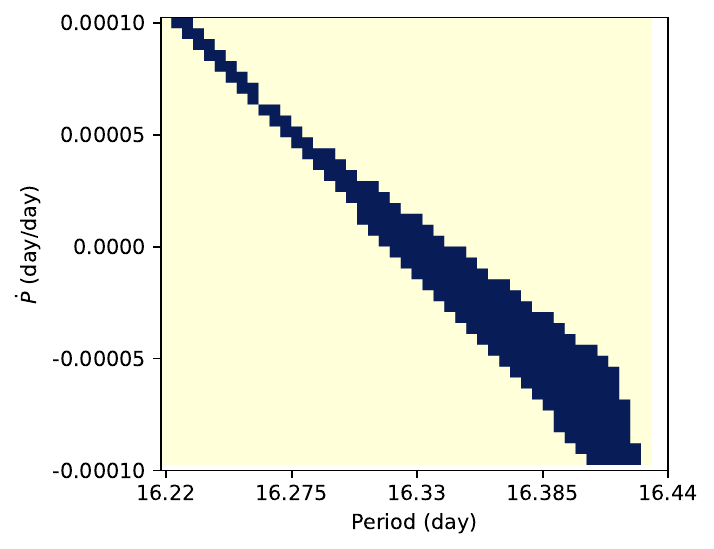}
	\caption{Constraints on the period derivative $\dot{P}$ and period $P$. The range of period $P$ is adopted from Table~\ref{table1}. The blue area is acceptable, which means that all of the subsequent 61 bursts can fall into the active window predicted by the earlier 61 bursts. The yellow area is not acceptable, which means that there are some bursts can not fall into the active window predicted by the earlier 61 bursts.}
	\label{fig_p}
\end{figure}

For the second method, we use the earlier 61 bursts from CHIME to establish an active window, and then verify if the subsequent 61 bursts fall within the active window. This method is similar to the one used to get the error bar of the period of FRB 20180916B \citep{2020Natur.582..351C}. We derive the $\dot{P}$ range for each period value, as shown in Fig.~\ref{fig_p}. The blue part is acceptable and the yellow area is unacceptable. Using this method, we get the $\dot{P}$ range for the best fitting results of the period in Table~\ref{table1}. The result is $-6\times10^{-5}< \dot{P} <2\times10^{-5}$ day day$^{-1}$ as listed in Table~\ref{table2}.

\begin{table}
	\centering
	\caption{The range of $\dot{P}$ of different intial periods for FRB 20180916B}
	\begin{tabular}{cc}
		\hline
		Period            &           $\dot{P}$ range (day  day$^{-1}$) \\
		\hline
		16.32     &   $0\sim2\times10^{-5}$    \\
		16.33     &   $-5\times10^{-6}\sim1\times10^{-5}$    \\
		16.34     &   $-2\times10^{-5}\sim5\times10^{-6}$    \\    
		16.35     &   $-3\times10^{-5}\sim-5\times10^{-6}$    \\
		16.38     &   $-6\times10^{-5}\sim-2.5\times10^{-5}$    \\
		\hline
	\end{tabular}
	\label{table2}
\end{table}

\begin{figure}
	\centering
	\includegraphics[width=0.5\textwidth]{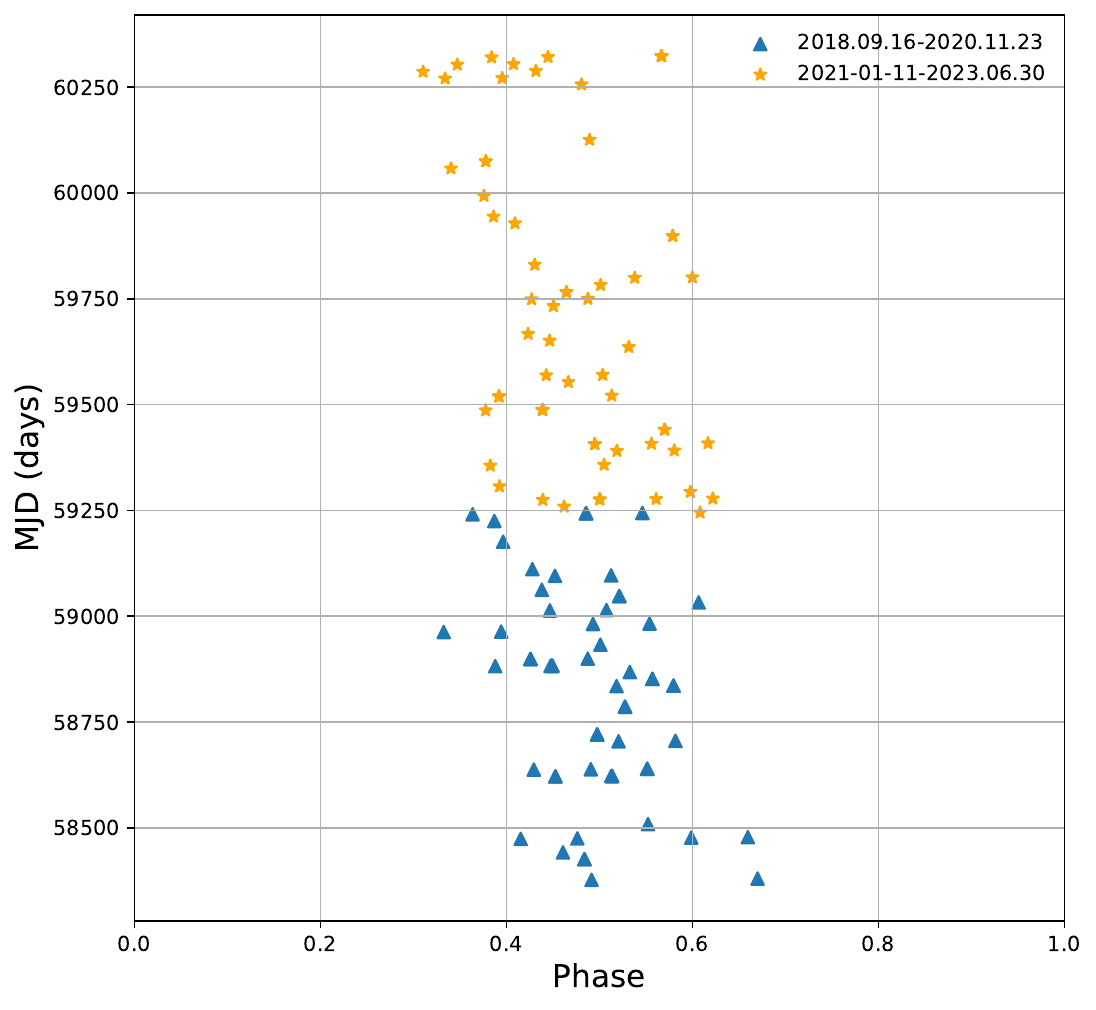}
	\includegraphics[width=0.5\textwidth]{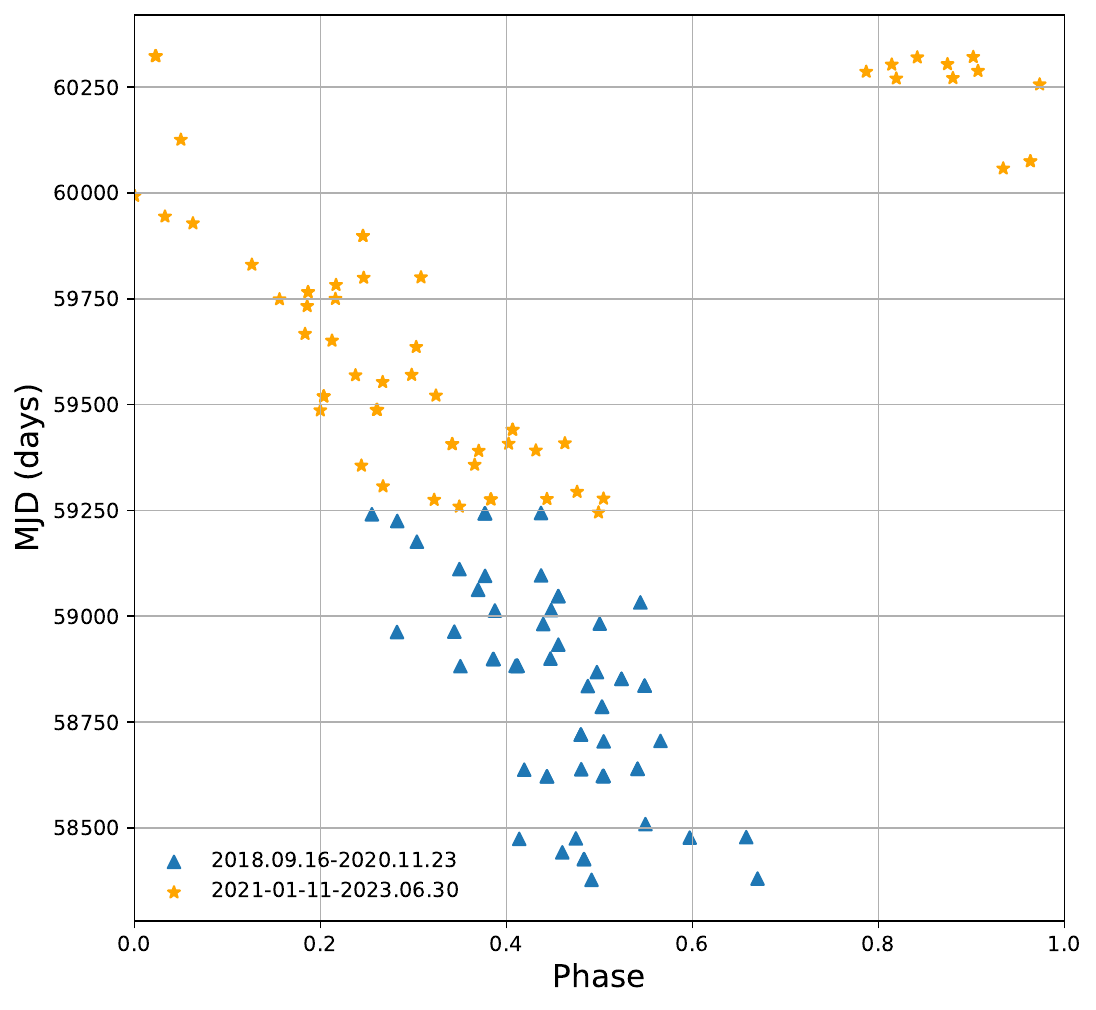}
	\caption{The relationship between MJD and phase for the bursts observed by CHIME after folding with two different $\dot{P}$. The bursts observed by CHIME are folded with Equation~(\ref{eq1}). The blue points represents the prior 56 bursts and the orange points represents the subsequent 55 bursts. The top panel shows an acceptable example with a period of 16.33 days and a $\dot{P}$ of $1.0\times10^{-5}$ day day$^{-1}$. The bottom panel shows an unacceptable example with a period of 16.33 days and a $\dot{P}$ of $8.6\times10^{-5}$ day day$^{-1}$. We can see in the bottom panel that the active window has a phase shift over time.}
	\label{fige}
\end{figure}

This method is based on the assumption that the active window remains the same over time. This assumption can be satisfied for three reasons. First, according to Table~\ref{table1} the $\dot{P}$ is relatively small, resulting in negligible changes in the period of the data over a given period of time. Second, for models of FRB 20180916B such as binary model or precession model, the active window will not change over time. Third, we believe that the active window provided by 61 bursts is sufficiently accurate. In order to mitigate the potential for slight inaccuracies, we extend the active window by 1.3 times its original width for verification. We give two examples in Fig.~\ref{fige} to help understand this method. The top panel of Fig.~\ref{fige} is an acceptable example and we can see that the bursts fall in a stable active window which is consistent with the prediction of the periodic model for FRB 20180916B. The bottom panel of Fig.~\ref{fige} is an unacceptable example. We can see the phase of the active window evolves over time which means the $\dot{P}$ is not suitable.

\section{Constraint on the periodic models}\label{sec3}
\subsection{The binary comb model}\label{sec3.1}
\begin{figure}
	\centering
	\includegraphics[width=0.6\columnwidth]{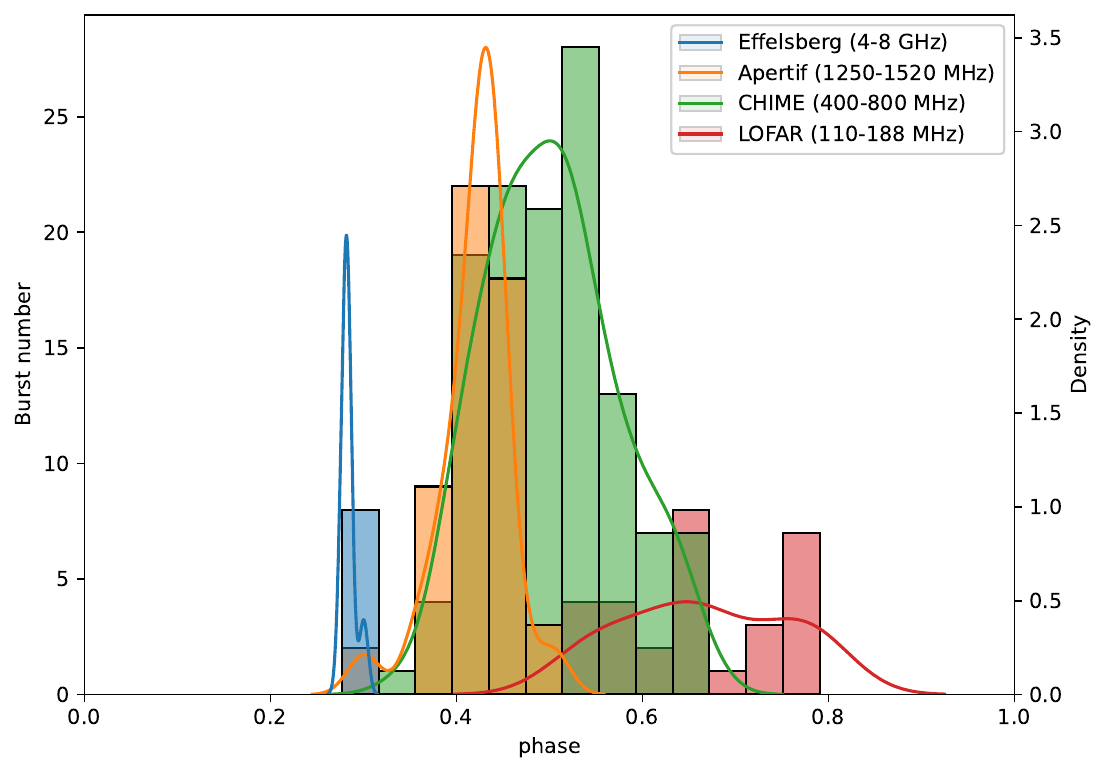}
	\caption{Chromatic periodic activity for different telescopes: Effelsberg \citep{2023MNRAS.524.3303B} (blue), Apertif \citep{2021Natur.596..505P} (orange), CHIME \citep{2019ApJ...885L..24C,2020Natur.582..351C,2023ApJ...950...12M} (green) and LOFAR \citep{2021Natur.596..505P,Pleunis2021} (red). The period used for the folding is 16.33 days, which is the best-fitting result with epoch-folding method. The colored lines are kernel density estimations. The active window is narrower and earlier at higher frequencies up to 8 GHz.}\label{fig_chro}
\end{figure}

With current observational data, we can test the periodic models for FRB 20180916B. First, the binary comb model is considered.
the binary comb model explains the 16-day periodicity by the orbital period, while the 5-day active window can be attributed to the free-free absorption of the companion star wind \citep{2020ApJ...893L..39L,2020ApJ...893L..26I,2021ApJ...920...54W}. We compile the bursts from recent observations detected by CHIME \citep{2019ApJ...885L..24C,2020Natur.582..351C,2023ApJ...950...12M}, LOFAR \citep{2024MNRAS.527.9872G} and Effelsberg radio telescope \citep{2023MNRAS.524.3303B}. The bursts detected by Effelsberg radio telescope can expand the frequency to 8 GHz. In Fig.~\ref{fig_chro}, we show the relationship between phase and active windows for bursts detected by different telescopes. The period used for folding is 16.33 days, which is the best result with the epoch-folding method. We can see that the active window is narrower and earlier at higher frequencies. Because the higher frequency FRBs are less likely obscured by the companion star wind, this model predicts a narrower active window in lower frequency, which contradicts the chromatic activity discussed above \citep{2021Natur.596..505P,Pleunis2021}. However, a revised binary model was proposed, which explains the chromatic activity of FRB 20180916B with the intrinsic emission mechanism \citep{2021ApJ...920...54W}. the binary comb model predicts that the DM of bursts varies within the active windows \citep{2021ApJ...920...54W,2020ApJ...893L..39L,2020ApJ...893L..26I}.
So, we perform the analysis of DM variance in different phases to test this prediction. We use all the bursts provided by CHIME and divide them into three bins with the same phase width. The value of $F$ distribution $F(2,119)=0.213$ and the $p$ value $p=0.81$ are derived. Here $F$ can be calculated by the following equation
\begin{equation}
    F = \frac{\sum_{i=1}^{k}(\overline{X_i}-\overline{X})^2/(k-1)}{\sum_{j=1}^{N}\sum_{i=1}^{k}(X_j-\overline{X_i})^2/(N-k)},
\end{equation}
where $k=3$ represents the number of factor and $N=122$ represents the number of bursts. $X$ represents the value of DM in a burst, $\overline{X_i}$ represents the mean value for factor $i$ and $\overline{X}$ represents the mean for all the bursts. Here, $k-1=2$ and $N-k=119$ in $F(2,119)$ represent the degrees of freedom for $F$ distribution and $p$ represents the possibility to accept the null hypothesis \citep{Sheskin1994}. In this case, the null hypothesis of this analysis is that the mean value of DM for the three phase bins is the same. The $p=0.81$ indicates that the phase is unrelated to the DM of bursts, which contradicts the prediction of this model. 

\subsection{The free precession model}\label{sec3.2}
Another explanation of the 16.3-day period is the free precession of a neutron star \citep{2020ApJ...895L..30L,2020ApJ...892L..15Z}. The five-day active window requires a fixed emission region, which can be explained by a dipole displaced from the neutron star center. The chromatic activity is attributed to the curvature radiation whose characteristic frequency is dependent on the altitude \citep{2021ApJ...909L..25L}. This model predicts an increase of the period of FRB 20180916B over time due to the dissipation of mechanical energy \citep{2021MNRAS.502.4664K,2022A&A...658A.163W}. Although there is no obvious increasing trend for the period of FRB 20180916B, the lifetime of free precession derived from the error bar of $\dot{P}$ is still in an acceptable range. However, there is another constraint on the internal magnetic field of the neutron star.

We consider a neutron star with free precession, where the deformation from the internal magnetic field $B_{\mathrm{int}}$ contributes to the precession. This phenomenon has been studied in previous work \citep{2020ApJ...892L..15Z}.
In this scenario, the precession period of the neutron star will increase as a result of the dissipation of mechanical energy. Specifically, there exists a relationship between the internal magnetic field $B_{\mathrm{int}}$ and the period derivative of precession $\dot{P}_{\mathrm{pre}}$ \citep{2021MNRAS.502.4664K,2022A&A...658A.163W}, which can be expressed as follows:
\begin{equation}\label{eq_free_precession}
	B_{\mathrm{int}} = \frac{G}{10\beta R^2_{\mathrm{ns}}\cos\theta}\sqrt{\frac{5\pi^2M^3_{\mathrm{ns}}}{6c^3P_{\mathrm{pre}}\dot{P}_{\mathrm{pre}}}}.
\end{equation}
Here, $\beta$ is a constant depending on the topology of the magnetic field and typically we have $|\beta|\ll 1$ \citep{2020ApJ...892L..15Z}. $\dot{P}_{\mathrm{pre}}$ is the period derivation of the free precession, $M_{\mathrm{ns}}$ is the mass of the neutron star which is set as $1.4$ $M_{\odot}$, $R_{\mathrm{ns}}$ is the radius of the neutron star which is set as $10$ km and $G$ is the gravitational constant.

\begin{figure}
	\centering
	\includegraphics[width=0.6\columnwidth]{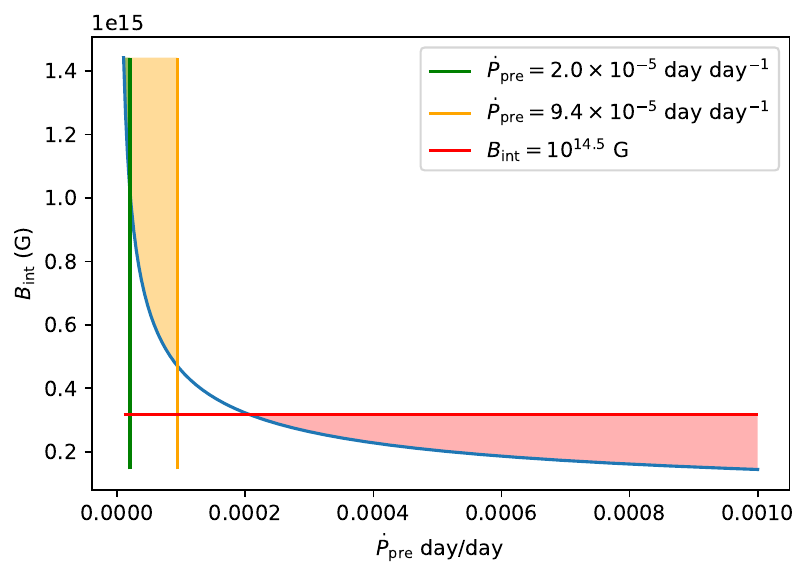}
	\caption{Constraint on the internal magnetic field $B_{\mathrm{int}}$ and period derivative $\dot{P}_{\mathrm{pre}}$ for the free precession model. The Blue line shows the relationship between $B_{\mathrm{int}}$ and $\dot{P}_{\mathrm{pre}}$ from Equation~(\ref{eq_free_precession}). The orange line represents the $\dot{P}_{\mathrm{pre}}=9.4\times10^{-5}$ day day$^{-1}$ and the orange area shows the acceptable value for $\dot{P}_{\mathrm{pre}}$. The green line represents the $\dot{P}_{\mathrm{pre}}=2.0\times10^{-5}$ day day$^{-1}$ and the green area shows the acceptable value for $\dot{P}_{\mathrm{pre}}$. The red line shows the $B_{\mathrm{int}}=10^{14.5}$ G and the red area represents the allowable region for the upper limit given by X-ray and $\gamma$-ray observation \citep{2020ApJ...893L..42T}.}
	\label{fig1}
\end{figure}

With this equation, we show the relationship between $B_{\mathrm{int}}$ and $\dot{P}_{\mathrm{pre}}$ in Fig.~\ref{fig1}. With the upper limit of $\dot{P}_{\mathrm{pre}}$ derived above, the lower limit of $B_{\mathrm{int}}$ can be given.
Astro-Rivelatore Gamma a Immagini Leggero (AGILE) and Swift observed the source of FRB 20180916B in X-ray and $\gamma$-ray bands \citep{2020ApJ...893L..42T}. They detected no extended X-ray and $\gamma$-ray emission, and gave an upper limit of the internal magnetic field \citep{2020ApJ...893L..42T,2022A&A...658A.163W}, estimated to be in the range of $10^{13.5}-10^{14.5}$G. 
Here, with the conservative value $\cos\theta<1$, $\beta<0.1$ and $\dot{P}_{\mathrm{pre}}<2.0\times10^{-5}$ day day$^{-1}$, 
we have $B_{\mathrm{int}}>1.02\times10^{15}$ G, shown in the green part of Fig.~\ref{fig1}. It well exceeds the upper limit of $B_{\mathrm{int}}=10^{14.5}$G given by the X-ray and $\gamma$-ray observation \citep{2020ApJ...893L..42T} as shown in the red part of Fig.~\ref{fig1}. So the free precession model cannot explain the active period.

\subsection{The forced precession model}\label{sec3.3}

\begin{figure}
	\centering
	\includegraphics[width=0.6\columnwidth]{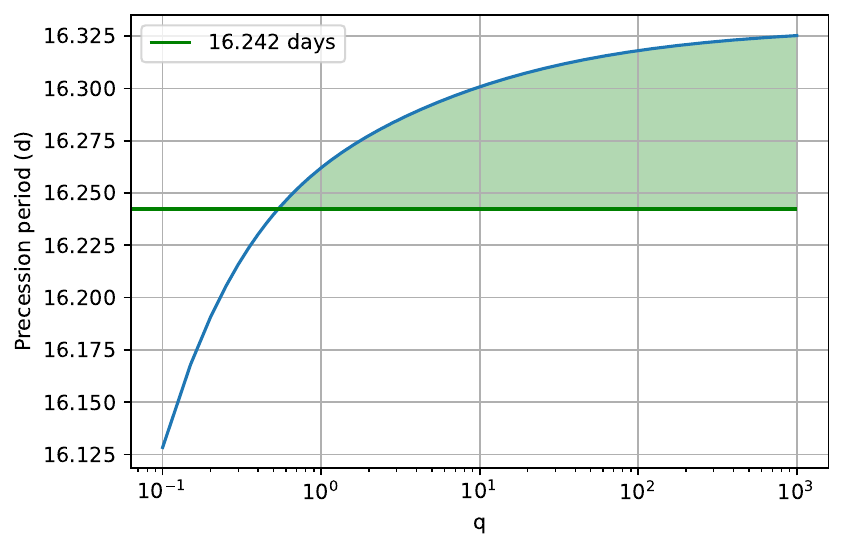}
	\caption{The relationship between mass ratio $q$ and the period after the evolution of four years for orbit-induced spin model. The initial period is set as 16.33 days which is the best fitting results. The blue line shows the period after the evolution of four years as a function of  mass ratio $q$. The green line shows the period of 16.242 days which is the temporal results of $\dot{P}_{\mathrm{pre}}=-6.0\times10^{-5}$ day day$^{-1}$. The green area are the acceptable range for mass ratio $q$.}
	\label{fig7}
\end{figure}

Forced precession of a neutron star is another model proposed to explain the period of FRB 20180916B. One of the forced precession models is the orbit-induced spin precession model \citep{2020ApJ...893L..31Y}, in which the FRB source possesses a precession due to the influence of a companion around it. 
In this scenario, the precession period is decreasing due to the radiation of gravitational waves. Assuming $q=\frac{M_1}{M_2}$ where $M_1$ is the mass of the companion and $M_2$ is the mass of the neutron star, we can give a constraint on $q$ with the relationship between mass ratio $q$ and $\dot{P}_{\mathrm{pre}}$ given by \cite{2020ApJ...893L..31Y}. 
Here, we assume the initial period as 16.33 days which is the best fitting result as discussed above and calculate the period after evolution of four years for different mass ratios $q$. Then by comparing with the $\dot{P}_{\mathrm{pre}}$ we get above, we have $q>0.55$ as shown in Fig.~\ref{fig7}.

For the fall-back disk model, the neutron star will precess when the rotation axis becomes misaligned with the normal axis of the fall-back disk \citep{2020RAA....20..142T}. 
With the equation given by \cite{2022A&A...658A.163W} and with $|\dot{P}_{\mathrm{pre}}|\le6.0\times10^{-5}$ day day$^{-1}$, we can get
\begin{equation}
	B_{\mathrm{int}}\ge1.00\times10^{15}\mathrm{G}|\sin\theta|\frac{0.1M_{\odot}}{M_{0}\cos\theta_{\mathrm{fb}}}\label{eq_fall},
\end{equation}
where $M_0$ is the total mass of the fall-back disk, $\theta$ is the angle between the rotational axis and the precession axis and $\theta_{\mathrm{fb}}$ is the angle between the rotational axis and the normal direction of a fall-back disk.
Since there are many uncertain parameters, the fall-back disk model can not be constrained by the internal magnetic field. 

\begin{figure}
	\centering
	\includegraphics[width=0.6\columnwidth]{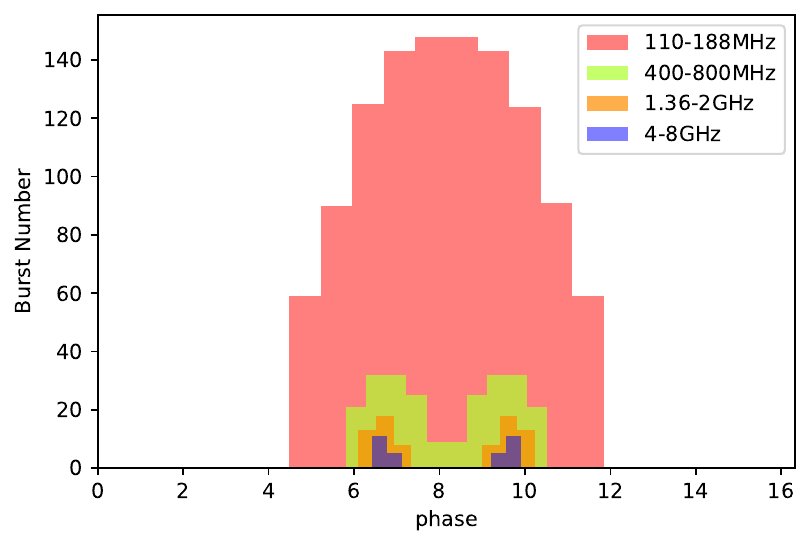}
	\caption{ The predicted burst numbers of forced precession model for different frequency bands: 110-188MHz (red), 400-800MHz (green), 1.36-2GHz (orange) and 4-8GHz (blue) under the assumption that the FRBs are emitted due to the curvature radiation. The frequency bands given in this figure are consistent with the frequency bands of LOFAR \citep{2021Natur.596..505P,Pleunis2021}, CHIME \citep{2019ApJ...885L..24C,2020Natur.582..351C,2023ApJ...950...12M}, Apertif \citep{2021Natur.596..505P} and Effelsberg \citep{2023MNRAS.524.3303B}.}
	\label{fig8}
\end{figure}

However, the phenomenon that the centers of the active windows for different frequency bands are in different phases is challenging in the context of the forced precession models \citep{2021ApJ...909L..25L}. Fig.~\ref{fig8} shows the prediction that there is no phase shift for different frequency bands. So the model is contradicted by the previous observations.

\subsection{The ultra-luminous X-ray binary model}\label{sec3.4}
Except the neutron star, FRBs can be also generated from an ultra-luminous X-ray binary, in which there are a compact object and a companion star undergoing sustained super-Eddington accretion \citep{Katz2020,2021ApJ...917...13S}. In this case, the 16-day period can be attributed to the precession of the compact object. The chromatic activity may result from the curvature of the quiescent jet cavity due to the motion of the disk winds driven by precession motion. However, in order to explain the extremely high brightness of FRBs, the system must undergo unstable mass transfer whose lifetime is about (5-100)$P_{\mathrm{orb}}$ \citep{2022ApJ...937....5S}. For a 16-day precession period, the $P_{\mathrm{orb}}$ is less than one day, which means the lifetime of the system is less than 100 days and this is much less than the active duration of FRB 20180916B. 
Another difficulty is the evolution of RM. This model predicts that the $|$RM$|$ of bursts decreases when the DM also decreases \citep{2022ApJ...937....5S}. However, the observed data shows a decreasing trend in the $|$RM$|$ of the bursts, while the DM of bursts remains constant. 

\subsection{The ultra-long rotation model}\label{sec3.5}
For the ultra-long rotation model, the period of FRB 20180916B is considered as the rotational period of a neutron star. The spin down of the rotation is enhanced due to the particle winds or a fallback accretion. So the neutron star can still possess strong magnetic field to power FRBs \citep{2020MNRAS.496.3390B}. The phase shift and different width of active window for different frequency bands can be explained by the curvature radiation of a displaced dipole consistent with the explanation of free precession above \citep{2021ApJ...909L..25L}. The period of rotation will keep stable \citep{2020MNRAS.496.3390B} which is consistent with the $\dot{P}$ derived from observation. Although it has been not found a neutron star with such a long rotational period, recently two observations reported their discovery of ultra-long period magnetars \citep{2023MNRAS.520.1872B} and radio transient \citep{2023Natur.619..487H}, which extended the population of radio transients with ultra-long period. PSR J0901-4046 was found to have a period of of 76 s \citep{Caleb2022} and GLEAM-X J162759.5 was argued to be a magnetar with a period of 1091 s \citep{Hurley-Walker2022}. Recently, a 21-min period radio transient GPMJ1839-10 was reported \citep{2023Natur.619..487H}. Therefore, the discovery of neutron stars with ultra-long rotational periods is promising in the future. 

With the help of the Hubble Space Telescope and 10.4 m Gran Telescope Canarias, an optical and infrared imaging as well as integral field  spectroscopy observations were presented on FRB 20180916B \citep{2021ApJ...908L..12T}. These observations suggest that the source of FRB 20180916B is 250 pc away from the nearest young stellar clump where there are lack of stars to create a magnetar \citep{2021ApJ...908L..12T}. If the neutron star is born in the star-forming region and travels to the place of FRB 20180916B, it will need 800 kyr to 7 Myr which is much greater than 10 kyr, which is typical for a highly energetic magnetar. However, there still exits the possibility to create a neutron star in the place of FRB 20180916B. Firstly, a compact-binary mergers or accretion-induced collapse can substitute for the core collapse of a massive star to create a neutron star \citep{Margalit2019,Wang2020}. In addition, a B star ejected from the dense stellar cluster can possess high velocities ($>30$ km s$^{-1}$) through binary interactions. With this velocity, the B star is able to move to the region of FRB 20180916B and produce the neutron star with a supernova explosion. Because of that, utilizing a neutron star as the source of FRB 20180916B is still possible.

\section{A self-consistent model for FRB 20180916B}\label{sec3.6}
\begin{figure*}
	\centering
	\includegraphics[width=0.8\textwidth]{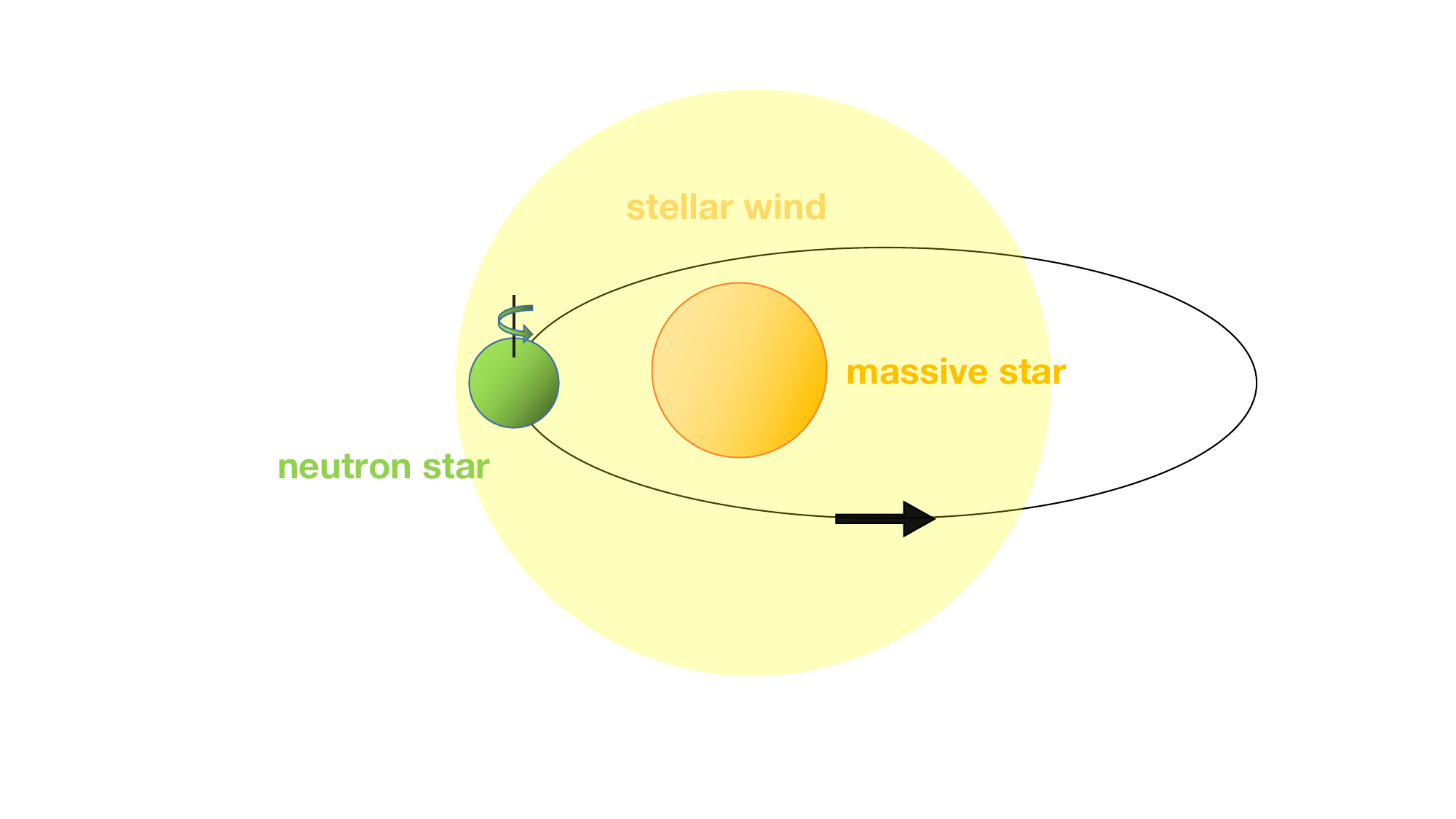}
	\caption{The Schematic diagram of the self-consistent model for FRB 20180916B. The green circle corresponds to the neutron star with a ultra-long rotation period. The orange circle represents the massive star with stellar wind shown as yellow part. The 16-day period is due to the ultra-long rotation of the neutron star and the variation of the RM is attributed to the orbital motion of this binary system.}
	\label{fig9}
\end{figure*}

Another crucial feature of FRB 20180916B is the increase of the RM with the stability of DM in the meanwhile. This phenomenon is difficult to explain in the context of the ultra-long rotation model. For isolated neutron stars, the value of RM is stable. Below, we propose a self-consistent model for FRB 20180916B as shown in Fig.~\ref{fig9}, which can accommodate the wealth of observational features for FRB 20180916B, especially the periodic activity and RM variation. 

\begin{figure}
	\centering
	\includegraphics[width=0.6\columnwidth]{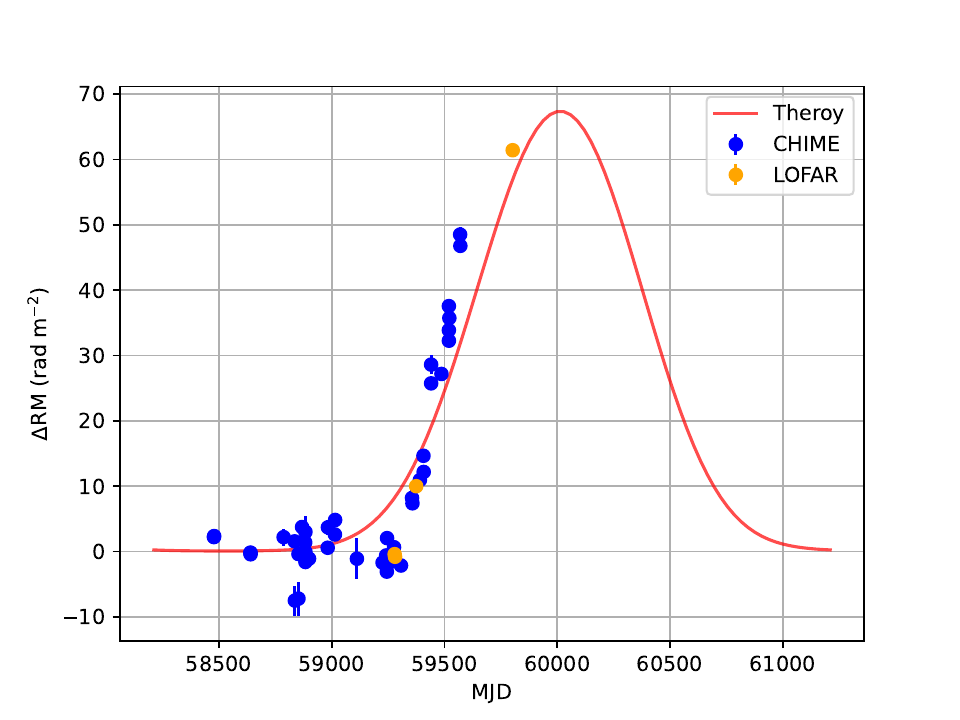}
	\includegraphics[width=0.6\columnwidth]{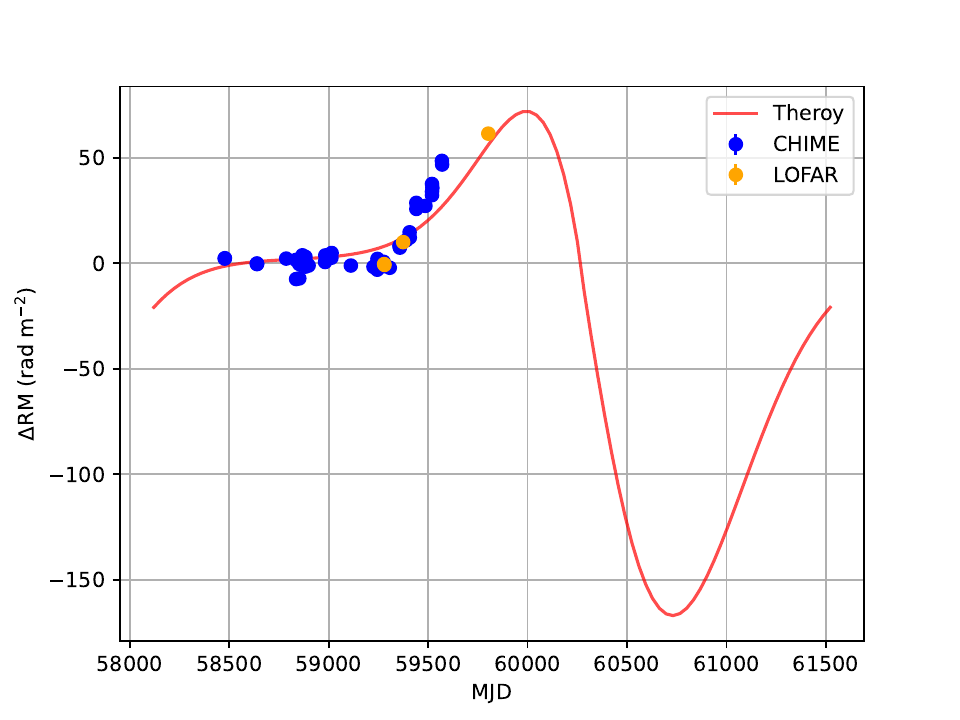}
	\caption{The RM variation of the self-consistent model. The y-axis $\Delta$RM=RM-RM$_{0}$ represents the variation of RM. Here, RM$_0=-115.8$ which is the mean value of the non-evolving phase \citep{2023ApJ...950...12M}. The blue and orange point shows the observational data from CHIME and LOFAR \citep{2023ApJ...950...12M,2024MNRAS.527.9872G}. The red line shows the results calculated with use of our model and fitting parameters. The top panel shows the RM variation of radial magnetic field and the bottom panel shows the the RM variation of toroidal magnetic field.}
	\label{fig10}
\end{figure}

We consider a massive binary containing a slowly rotational neutron star and a massive star with large mass loss. Radio bursts are possibly generated in the magnetosphere of the neutron star through curvature radiation \citep{Yang2018,Lu2020}. Different from the binary comb model discussed in Section~\ref{sec3.1} which use the orbital period to explain the period of FRB 20180916B, the period of our model can be caused by the ultra-long rotation of the neutron star, and the variation of RM is attributed to the mass loss of the massive star. Within this model, RM variations are caused by the interaction between the radio signal and the magnetized decretion disk or stellar wind during periastron passage of the neutron star. An interacting binary system featuring a magnetar and high-mass stellar star had been proposed to explain the RM variations of FRB 20201124A \citep{Wang2022} and FRB 20190520B \citep{Wang2022,Anna-Thomas2023}. In this scenario, the variation of RM is unrelated to the period \citep{2023ApJ...942..102Z}. For a large space of model parameters, this model can explain the increase of RM and the stability of DM for FRB 20180916B \citep{2023ApJ...942..102Z}. 
In Fig.~\ref{fig10}, we fit the variation of the RM with use of the data observed by CHIME \citep{2023ApJ...950...12M} and LOFAR \citep{2024MNRAS.527.9872G}. Since the variation of RM spans over 4 years, the orbital period must be very long. This results in a significant large distance between the neutron star and the massive star. Because of that, our model is different from the high mass X-ray binary since it is difficult for the neutron star to accrete the stellar wind and the luminosity of X-ray is about $1.16 \times 10^{30}$ erg s$^{-1}$ \citep{2016A&A...589A.102B}. In this case, we consider the magnetic field of the stellar wind is radial or toroidal
\begin{equation}
	B_{\mathrm{s}}=B_0(\frac{R_{\mathrm{A}}}{R_{\star}})^{-3}(r/R_{\mathrm{A}})^{-\alpha},
\end{equation}
where $R_{\mathrm{A}}$ is the the Alfvén radius, $R_{\star}$ is the radius of the star, $B_0$ is the magnetic field strength at $R_{\star}$ and $r$ is the radial distance. Here, $\alpha$ is a constant that $\alpha=1$ represents a radial field and $\alpha=2$ represents a toroidal field when the OB star possesses a fast rotation velocity \citep{1992ApJ...395..575U}.
Otherwise, for the massive star with $30 M_{\odot}$, its electron number density of the stellar wind is given by 
\begin{equation}
	n_{\mathrm{w}}(r)=n_{\mathrm{w,0}}(\frac{r}{R_{\star}})^{-2}
\end{equation}
where $n_{\mathrm{w,0}}$ is the number density of electron at the surface of the massive star. Assuming a constant mass-loss rate $\dot{M}$ and a constant wind velocity, the $n_{\mathrm{w,0}}$ can be written as $n_{\mathrm{w,0}}=\dot{M}/4\pi R_{\star}^2v_{\mathrm{w}}\mu_{\mathrm{i}} m_{\mathrm{p}}$, where the mass-loss rate is set as the typical value of Be stars which is about $10^{-9}$ $M_{\odot}$yr$^{-1}$, $\mu_{\mathrm{i}}$ is the mean ion molecular weight set as 1.29 \citep{2013A&ARv..21...64D} and $m_{\mathrm{p}}$ is the mass of the protons. We adopt the method outlined by \cite{2023ApJ...942..102Z} to fit the observational data. Other typical values are presented in Table~\ref{table3}.
\begin{table*}
	\centering
	\caption{The physical parameters used to fit the RM variation for the radial magnetic field and the toroidal magnetic field models.}
	\label{table3}
	\begin{tabular}{ccccc}
		\hline
		Parameter  & Symbol  & Radial & Toroidal  & Reference\\
		\hline
		Orbital period & $P$ & 3000 d & 3400 d & \\
		Eccentricity & $e$ & 0.7 & 0.6 & \\
		Surface magnetic field & $B_0$ & 40 G &3.9 G &  \\
		Inclination angle of observers & $i_{\mathrm{o}}$ & 26$^{\circ}$ & 20$^{\circ}$ & \\
		True anomaly angle of observers & $\phi_{\mathrm{o}}$ & 132$^{\circ}$ & 132$^{\circ}$ & \\
		\hline
		Mass of star & $M_{\star}$ & 30 $M_{\odot}$& 30 $M_{\odot}$ & (1)\\
		Radius of star & $R_{\star}$ & 10 $R_{\odot}$& 10 $R_{\odot}$ & (1,2)\\
		Effective temperature of star & $T_{\mathrm{eff}}$ & $3\times10^{4}$ K & $3\times10^{4}$ K & (1)\\
		\hline
		Mass-loss rate & $\dot{M}$ & $2\times10^{-9}$ $M_{\odot}$yr$^{-1}$ & $10^{-9}$ $M_{\odot}$yr$^{-1}$ & (3)\\
		Wind velocity & $v_{\mathrm{w}}$ & $3\times10^{8}$ cm s$^{-1}$ & $3\times10^{8}$ cm s$^{-1}$ & (3)\\
		Wind temperature slope & $\beta_{\mathrm{w}}$ & $2/3$ & $2/3$ & (4,5)\\
		\hline
		Mass of pulsar & $m$ & 1.4 $M_{\odot}$ & 1.4 $M_{\odot}$ & \\
		Spin-down luminosity & $L_{\mathrm{sd}}$ & $10^{36}$ erg s$^{-1}$ & $2\times10^{35}$ erg s$^{-1}$ & (6,7)
	\end{tabular}
	\begin{tabular}{c}
		\textbf{References}: (1) \cite{2011ApJ...732L..11N} (2) \cite{2005MNRAS.364..899C} (3) \cite{1981ApJ...251..139S} (4) \cite{1993ApJ...406..638K} (5) \cite{2005ARep...49..295B} \\
		(6) \cite{1995ApJ...445L.137M} (7) \cite{2009ApJ...705....1C}
	\end{tabular}
\end{table*}

The RM variation of the fitting results are shown in Fig.~\ref{fig10} and the fitting parameters are shown in the first line of Table~\ref{table3}. For the radial magnetic field, the orbital period $P=3000$ d, the eccentricity $e=0.7$, the surface magnetic field of the massive star $B_0=40$ G. For the toroidal magnetic field, the orbital period $P=3400$ d, the eccentricity $e=0.6$, the surface magnetic field of the massive star $B_0=3.9$ G. The DM variations for both cases are also calculated to be less than $0.15$ pc cm$^{-3}$ which is significantly lower than the error of DM observation. In our model, the variation of RM is periodic which can be tested by future observations. 

Besides FRB 20180916B, there are also many repeating FRBs with large variation of RM \citep{Mckinven2023}. However, for an isolated neutron star, its pulsar wind, pulsar wind nebula and flares have a very small contribution to the RM and its variation \citep{2023MNRAS.520.2039Y}. Otherwise, while isolated young magnetars can generate FRBs \citep{CHIME/FRBCollaboration2020,Bochenek2020}, their short magnetic activity timescale is difficult to explain the high FRB rate \citep{2016ApJ...833..189L,Wang2020}. Both the two factors suggest that the binary system may have a significant advantage in explaining the observational properties of FRBs. This is because on the one hand, the stellar wind can produce variation of DM and RM. On the other hand, the accretion of the stellar wind can help the to produce FRBs especially for old magnetars. Initially, accretion from a companion could trigger the glitch of the magnetar. Such kind of a glitches have been discovered in accretion-powered pulsars by \cite{2017MNRAS.471.4982S} and \cite{2004ApJ...613.1164G} with a sudden frequency change of $\Delta v=1.28(5)\times10^{-6}$ Hz and $\Delta v=1.8\times10^{-6}$ Hz. These glitches are similar to the giant glitches of SGR J1935+2154 occurring before FRB-like bursts \citep{2024Natur.626..500H,2024RAA....24a5016G}. Such glitches could alter the core magnetic field during the relaxation phase of the glitch, which will eventually trigger the movement of the crust and produce FRBs through the crust quakes \citep{Wang2018,Suvorov2019}. Otherwise, the accretion of the Be star disk or stellar wind can also interact with the magnetosphere of the neutron star and change the spin period of the neutron star \citep{Li2021}. This results in the variation of the centrifugal force of the neutron star and could change the stress of the crust. After the stress of the crust reaches a critical value, the starquakes will be induced which could further produce FRBs \citep{Li2021,Lu2020}. For the accretion of Be star disk \citep{Li2021}, this model requires the neutron star is inside the disk of the Be star. However, the separation between the neutron star and the Be star is about 2700 $R_{\odot}$ for the period of 3000 d , which is much larger than the radius of the disk (about 100 $R_{\odot}$) \citep{2013A&ARv..21...69R}. For the accretion of the stellar wind, \cite{2016A&A...589A.102B} provides a criteria to judge the impact of the stellar wind, which is determined by three radii, i.e., the accretion radius, the magnetospheric radius and the corotation radius. The accretion radius represents the distance from the neutron star at which the stellar wind is captured by gravity
\begin{equation}
	R_{\mathrm{a}}=2GM_{\mathrm{NS}}/v^2_{\mathrm{rel}},
\end{equation}
where $v^2_{\mathrm{rel}}$ is the relative velocity of the neutron star relative to the stellar wind. The magnetospheric radius is the distance from the neutron star, at which the magnetic pressure balances the ram pressure of stellar wind,
\begin{equation}
	R_{\mathrm{M}}=1.3\times10^{10}\rho^{-1/6}_{-12}v^{-1/3}_{8}\mu^{1/3}_{33} \rm cm.
\end{equation}
Here, $\mu^{1/3}_{33}=1/2B_{\mathrm{d}} R_{\mathrm{ns}}^3/10^{33}$ G cm$^3$ and $\rho_{-12}=n_{\mathrm{w}}\mu m_{\mathrm{p}}/10^{-12}$ g cm$^{-3}$. The corotation radius represents the distance where the Keplerian angular velocity is equal to the spin rotation of the neutron star,
\begin{equation}
	R_{\mathrm{co}}=1.7\times10^{10}(P_{\mathrm{r}}/10^3 \mathrm{s})^{2/3},
\end{equation}
where the $P_{\mathrm{r}}$ is the rotation period of the neutron star. For FRB 20180916B, using the fitting parameters, we calculate the three radii and find that $R_{\mathrm{co}}>R_{\mathrm{M}}>R_{\mathrm{a}}$. In this scenario, the stellar wind material is prevented from accreting by the magnetic field but the material passing along the magnetospheric boundary can accrete through the Kelvin-Helmholtz instability but in our model. Since the distance between the star and the neutron star is too far, the density of the stellar wind near the neutron star is too low to change the spin period of the neutron star. For FRB 20180916B, we prefer the low-twist magnetar model proposed by \cite{2019ApJ...879....4W} and \cite{2020MNRAS.496.3390B}. However, for FRB sources whose DM variation is not very small, we can consider a short orbital period and highly structured stellar winds with large velocity and density grandients. For a binary system with a 30-day orbital period, $e=0.7$ and stellar mass loss rate of $10^{-8}$ $M_{\odot}$yr$^{-1}$, we find $R_{\mathrm{a}}>R_{\mathrm{M}}$ near the periastron. In this scenario, the magnetic field fails to obstruct stellar wind accretion, potentially greatly enhancing accretion efficiency \citep{2016A&A...589A.102B,Li2021}. The waiting time of the two starquakes induced solely by the accretion of the stellar wind is about $10^{2}$ days \cite{Li2021}, which demonstrates that the accretion of the stellar wind could potentially create an ideal environment for FRB production by triggering giant glitches or starquakes. Therefore, binary systems with a neutron star and a massive star not only can provide a RM and DM variation but also is a well environment for producing FRBs by accretion of the stellar wind or stellar disk.






\section{Conclusions}\label{sec4}
In this work, we derive the period and $\dot{P}$ with the bursts detected by CHIME and LOFAR in Section~\ref{sec2} and find the period keep stable over time. Then, we use the observational data to constrain the periodic model for FRB 20180916B. We find the ultra-long rotation model appears to be the best-fit periodic model for FRB 20180916B. However, the previous ultra-long rotation model considers an isolated neutron star which contradicts the recent observation of RM. Here we propose a self-consistent model, which is a massive binary containing a slowly rotational neutron star and a massive star with large mass loss. This modified model can naturally accommodate the wealth of observational features for FRB 20180916B and the accretion of the stellar wind creates an ideal environment for producing FRBs. Moreover, the RM variation of this model is periodic, which can be tested by future observations.

\section*{Acknowledgements}
We thank the anonymous referee for helpful comments. This work was supported by the National Natural Science Foundation of China (grant No. 12273009), the National SKA Program of China (grant No. 2022SKA0130100), and the China Manned Spaced Project (CMS-CSST-2021-A12). We acknowledge use of the CHIME/FRB Public Database, provided at https://www.chime-frb.ca/ by the CHIME/FRB Collaboration.

\bibliographystyle{aasjournal}
\bibliography{example} 








\label{lastpage}
\end{document}